# Giant molecular clouds: s

**Clare Dobbs discusses star formation in galaxies and the sites of star formation: giant molecular clouds.**

Stars are forming in our galaxy at a rate of between 1 and 4 solar masses of stars per year. In contrast to elliptical galaxies, which are largely devoid of star formation, star formation is still going on in spiral galaxies because of their reservoirs of molecular gas, the fuel for new stars. The discs of spiral galaxies are comprised not only of stars as we clearly see from Earth, but also gas (the interstellar medium, ISM). This is where this gas accumulates into cold, dense, molecular regions known as molecular clouds, in which new stars are formed. Most star formation occurs in massive molecular clouds, known as giant molecular clouds (GMCs). However, while we have a good understanding of how individual stars form, there is less consensus on how their natal clouds of gas accumulate, how long these clouds last, how star formation progresses over their lifetime, and indeed how star formation has progressed over the lifetime of the Milky Way.

What we do know about star formation in nearby galaxies tells us that the rate of star formation is surprisingly low, but we do not know why. In order to do this we need to study the evolution of the gas, and how it is turned into stars.

Understanding the formation and evolution of GMCs, though, is a formidable problem. One immediate challenge is the vast range in scales between galaxies and protostellar discs, from ~10 kpc across down to ~$10^{-3}$ pc. Another difficulty is the complex physics involved: gravity, magnetic fields, thermodynamics, turbulence and stellar feedback all play roles. The ISM itself is a multiphase medium of atomic, molecular and ionized hydrogen spanning a range of temperatures from 10 K to >$10^8$ K, and many orders of magnitude in density.

Theoretically there are two main ideas of how GMCs form: gravitational instabilities, and the agglomeration of smaller clouds. Gravitational instabilities will lead to the collapse of gas on unstable wavelengths, as originally described by Jeans in the context of uniform spheres of gas. The stability of the gas to gravitational collapse in a galaxy depends on the mass of the gas in the disc, the gas temperature and the rotation of the galaxy (Toomre 1964, Goldreich and Lynden-Bell 1965). GMC formation has also been proposed by collisions between, or agglomeration of, smaller clouds of gas (atomic

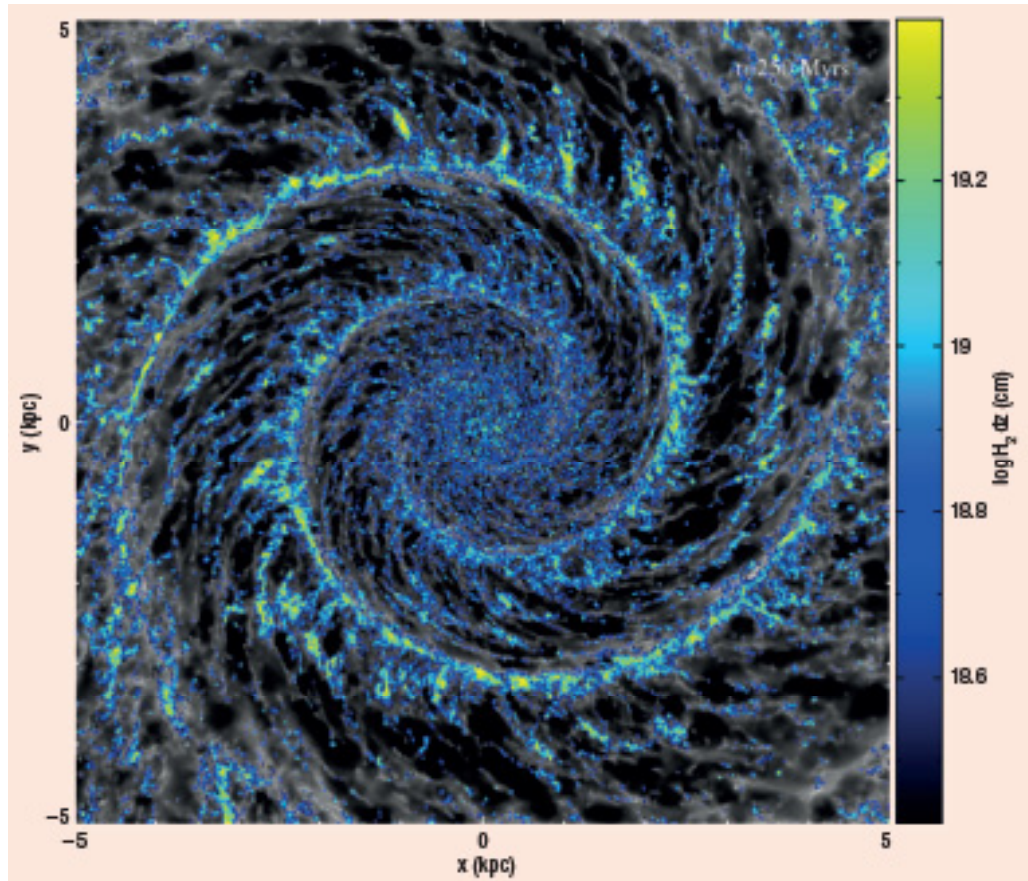

1 (a): Simulation of a galaxy disc showing the distribution of gas after 250 Myr (from Dobbs and Pringle 2013). T agglomeration of smaller clouds, and self-gravity. (b): Hubble map of the inner region of the Whirlpool Galaxy or molecular). In this case, continued mergers of small clouds allow the build up of massive GMCs. There are also other suggested mechanisms for GMC formation, for example Parker instabilities (although these do not produce such large density enhancements; Kim et al. 2002), as well as colliding flows from turbulence, supernovae, or stellar winds, though these probably produce only smaller clouds.

As well as studying how GMCs form, we also need to know how these clouds are dispersed, or destroyed. The gas in GMCs could all be turned into stars, but this would lead to much higher star formation rates than observed, unless the clouds are extremely long-lived. It is more likely that processes such as stellar feedback (supernovae, winds, radiation pressure), galactic shear and turbulence cause the clouds to disperse, leading to the termination or reduction of star formation.

Because of the complex physics involved,

> **"The gas in GMCs could all be turned into stars, but this would lead to much higher star formation rates"**

numerical simulations of galaxies are ideally required to follow GMC formation and evolution. These were not feasible until around 2000, and those by Wada and Norman (1999) are among the first. The calculations need to model the gas in the galaxies, so they treat the gas as a fluid and solve the fluid dynamics equations. Some simulations also explicitly include the stellar disc, with the stars obviously subject only to gravitational forces. Due to the difficulty in modelling a whole galaxy, however, other simulations do not explicitly model the dark matter halo and stars. Rather, these are assumed to be a fixed (or rotating) gravitational potential. Then the force due to the potential ($-\nabla\phi$) can be simply added to the momentum equation. Even with the progress in computational resources, and any simplifications, calculations still typically take months to complete on supercomputers.

Simulating isolated galaxies means that it is

# tar factories of the galaxy

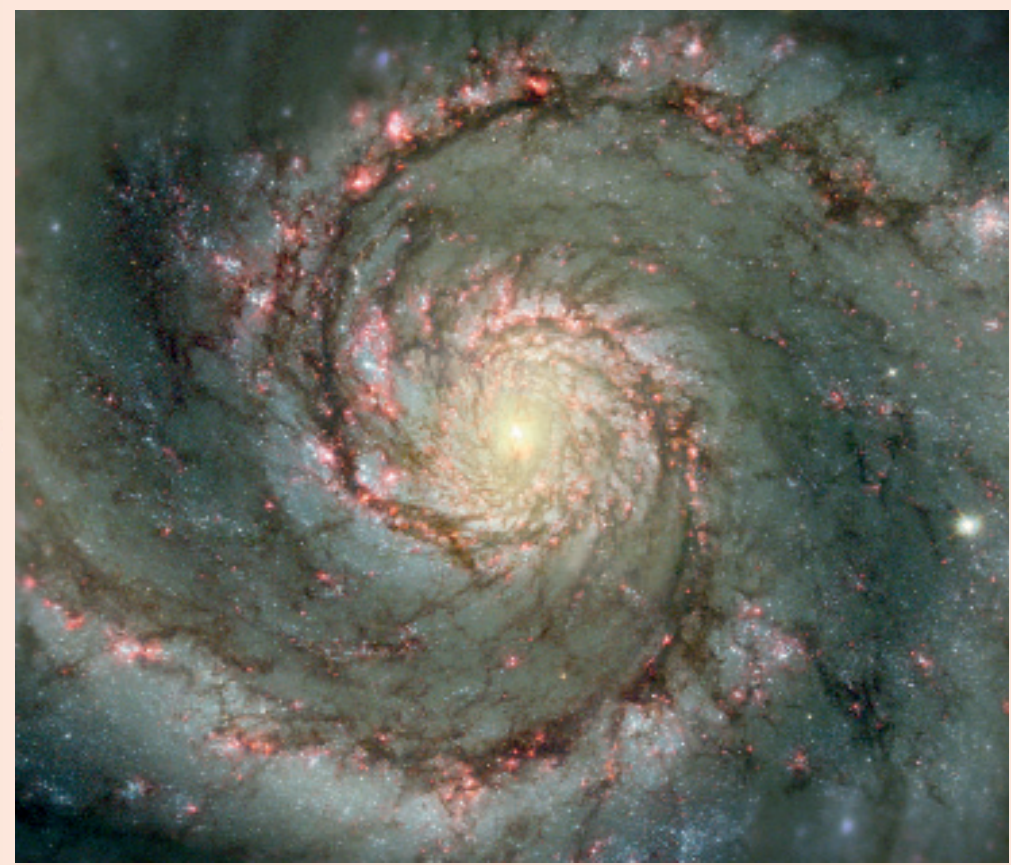

he colour scale represents the amount of molecular gas. The gas arranges into dense clouds and spurs by the (M51).

possible to reach much higher resolution compared to cosmological simulations, but how the galaxy forms, galaxy mergers and accretion of gas onto the galaxy are neglected. However, generally the timescales (hundreds of Myrs) of the simulations are quite short, so neglecting these processes is reasonable.

Spiral galaxies generally fall into two categories: grand design galaxies and flocculent galaxies. Flocculent galaxies have many transient, short spiral arms arising from local gravitational instabilities in the stars and/or gas, so require including both stars and gas in the simulations. Grand design spiral galaxies, which are typically symmetric two-armed galaxies, are modelled most easily (though not necessarily most realistically) by adopting a spiral perturbation of an underlying stellar potential.

### Evolution of GMCs

Figure 1a shows a hydrodynamic simulation from Dobbs and Pringle (2013), which includes a rigid spiral potential, gas thermodynamics, chemistry of $H_2$ and CO formation, self-gravity and stellar feedback (which represents the effects of supernovae and winds from massive stars). The figure shows the simulation at a time of 250 Myr. Magnetic fields are not included. The simulation is performed using smoothed particle hydrodynamics (SPH), a Lagrangian fluids code (this particular code is sphNG; Benz *et al.* 1990, Bate 1995, Price and Monaghan 2007). The simulation uses 8 million particles, giving a particle mass of $312.5 M_\odot$.

The implementation of stellar feedback to numerical simulations varies between simulations, and involves making assumptions about star formation at resolutions the simulations cannot reach, so some discussion is merited here of how this physics is added. Stellar feedback is inserted once a particle (or for grid codes, a grid cell) exceeds a given density threshold, at which point stars are assumed to form (see Dobbs *et al.* 2011). Stellar feedback represents the processes undergone by stars that influence the surrounding gas. Massive stars have disproportionate effects on the ISM – they have short lifetimes after which they undergo supernovae explosions, drive stellar winds, exert a large radiation pressure and ionize the surrounding gas. Feedback is inserted into the numerical simulations as kinetic and/or thermal energy into the surrounding gas (in this particular example, energy is added as a combination of kinetic and thermal). In this calculation, the amount of energy added corresponds to one supernova for every $160 M_\odot$ of stars formed. Unlike many numerical simulations of galaxies, though, the prescription used for the simulation shown in figure 1a does not adopt the empirical Schmidt–Kennicutt relation to assign star formation rates based on the density; rather the star formation rate is an output of the simulations. However, we still need to use an unknown efficiency parameter, because we do not have the resolution to model individual stars forming; we can only assume some fraction of the gas we resolve in our models forms stars. The efficiency parameter can be related to observational estimates of the star formation efficiency, and increasing or decreasing this parameter leads to more or less energy inserted in each star formation event.

The black and white colour scale on figure 1a shows the total column density of the gas. Regions that are molecular hydrogen are highlighted by the yellow and blue colour scale (the scale represents the molecular gas fraction integrated through the disc). The simulated galaxy is characterized by dense clumps along the spiral arms. These correspond to GMCs and are where most of the star formation occurs. The structure is not dissimilar to (false colour) Hubble images of observed galaxies, e.g. M51, M83. Figure 1b shows a Hubble image of the inner part of M51. In this image, the dense gas corresponds to the dark regions, while the red parts are regions of star formation.

The formation of the clouds in the simulation shown in figure 1a is likely to arise from a combination of agglomeration of smaller clouds and self-gravity. Self-gravity increases interactions between clouds, gas accretion onto clouds, and gravitational instabilities in the ISM. Dobbs (2008) showed that, generally, GMC formation is not simply confined to one process. For low gas surface densities (<$10 M_\odot pc^{-2}$), formation is dominated by agglomeration, whereas for higher surface densities (>$10 M_\odot pc^{-2}$), self-gravity starts to dominate. The surface density of the calculation shown in figure 1a is $8 M_\odot pc^{-2}$ (the surface density of the Milky Way is ~$10 M_\odot pc^{-2}$), so GMCs are predominantly formed by agglomeration, although self-gravity still has an effect on cloud structure. One earlier criticism of the formation of GMCs by agglomeration was the timescale needed to grow a massive cloud in this way (Blitz and Shu 1980).

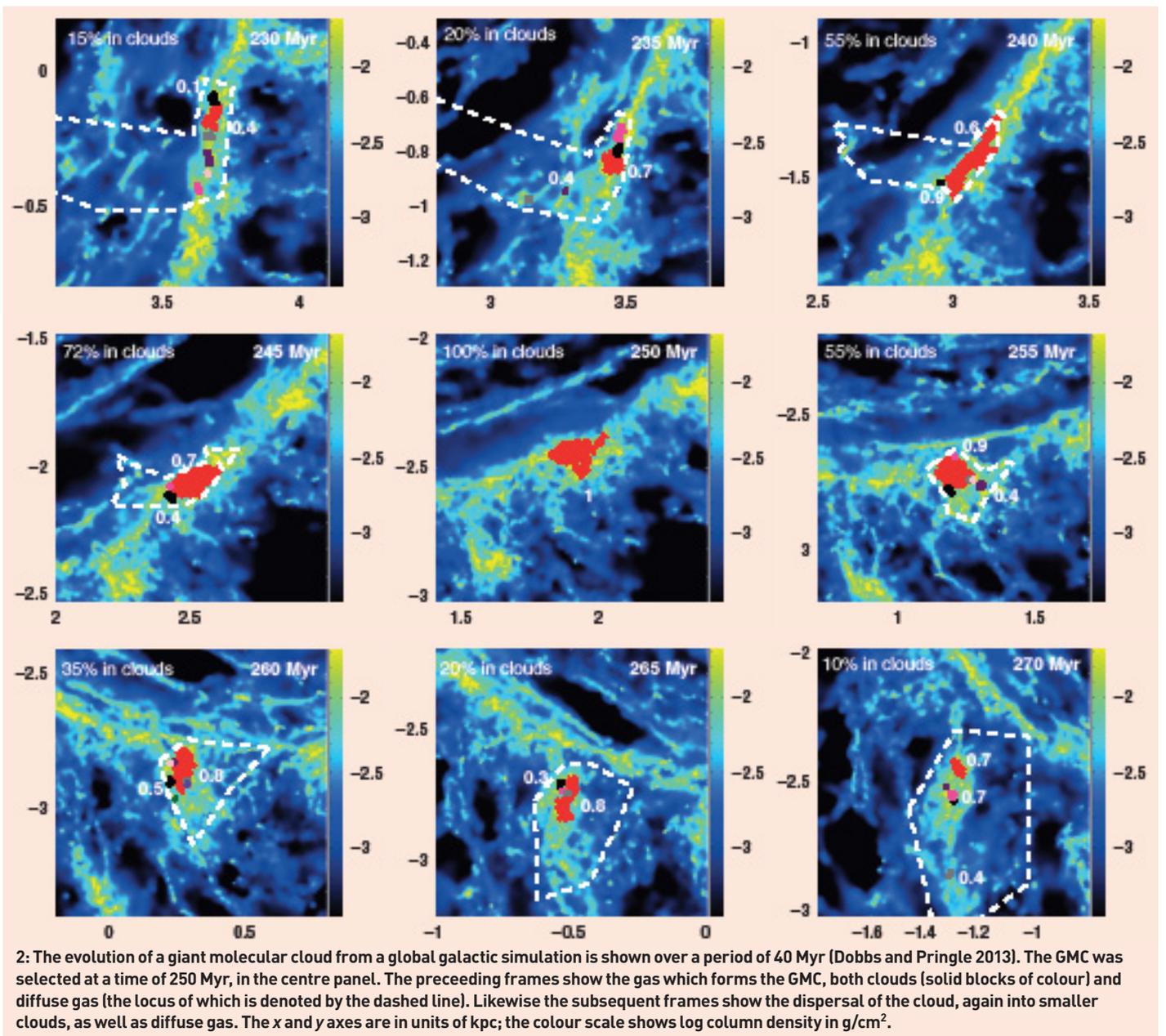

2: The evolution of a giant molecular cloud from a global galactic simulation is shown over a period of 40 Myr (Dobbs and Pringle 2013). The GMC was selected at a time of 250 Myr, in the centre panel. The preceeding frames show the gas which forms the GMC, both clouds (solid blocks of colour) and diffuse gas (the locus of which is denoted by the dashed line). Likewise the subsequent frames show the dispersal of the cloud, again into smaller clouds, as well as diffuse gas. The x and y axes are in units of kpc; the colour scale shows log column density in g/cm².

However, this problem is largely avoided here, because as the gas passes through the spiral arms, there are many more interactions and collisions between clouds allowing them to grow in a short space of time. Away from the spiral arms, or in the absence of spiral arms, cloud growth is much less efficient.

Also apparent in figure 1 are many long, thin interarm spokes, or spurs. These features form as a result of the shearing of dense molecular clouds along the arms. During their passage along the arms, the molecular clouds are not subject to strong shear, and therefore remain intact. However, when they move out of the arms into the interarm region, they become subject to shear as a result of the differential rotation of the galaxy. These features have been seen in several other simulations of grand design galaxies (Wada and Koda 2004, Shetty and Ostriker 2006) and numerous actual galaxies (LaVigne et al. 2006), most notably M51.

Between the spurs lie low-density regions, or holes, some of which have been accentuated by stellar feedback. Such holes are often seen in observed galaxies, particularly in H I maps, and the most massive, supershells, are assumed to be the result of multiple supernovae.

Figure 2 shows the detailed evolution of a cloud from this simulation, over a period of 40 Myr. The cloud has a mass of $2 \times 10^6 M_\odot$, and is situated at a galactic radius of 3 kpc. The cloud was selected at a time of 250 Myr (centre frame), then the preceeding and subsequent frames show the evolution before and after the cloud is selected. The cloud forms from a mixture of smaller clouds, shown as solid blocks of colour, and ambient ISM, and similarly disperses into small clouds and ambient ISM. The cloud disrupts by a combination of shear and stellar feedback. The effect of shear can be seen in figure 2: the cloud becomes more elongated, breaking up but with the smaller clouds still situated along a spur. Shear tends to act over larger scales, whereas stellar feedback tends to act over smaller scales and is important for breaking up gravitationally bound clouds, or gravitationally bound clumps within clouds. The timescale over which there is an obvious, single, massive cloud present, is around 25 Myr (from ~240–265 Myr). This is a rough measure of the lifetime of the cloud. The lifetimes of clouds in the simulations are generally up to around 30 Myr, but the lower mass, ~$10^5 M_\odot$ clouds tend to have shorter lifetimes of <10 Myr. Over their lifetime, the GMCs typically convert a few per cent of their mass into stars, a figure in agreement with other observations and other theoretical work.

### Properties of GMCs

One way of determining how well the simulations compare with real galaxies is to compute the properties of molecular clouds. It is important to test how well simulations reproduce

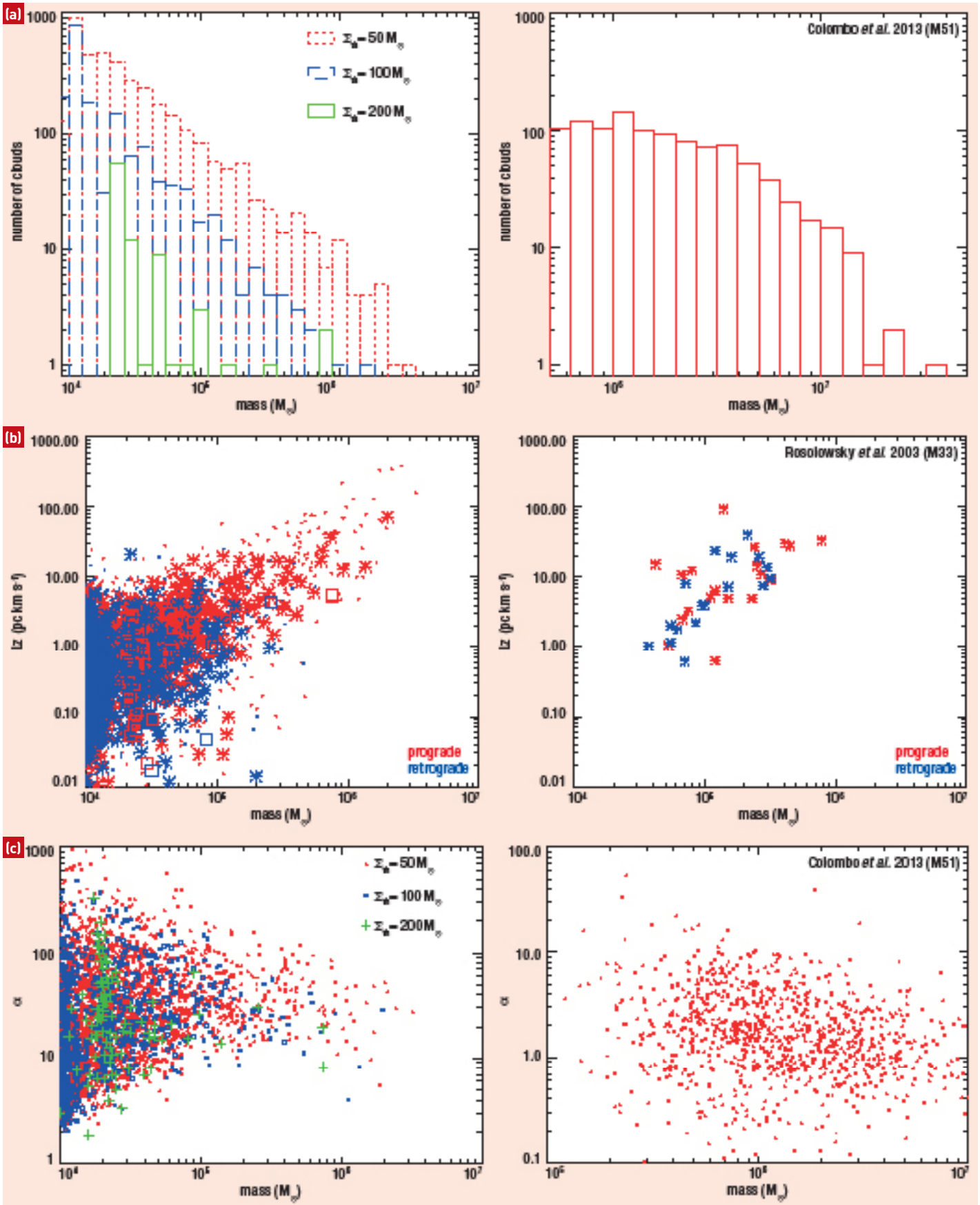

3: The properties of GMCs are shown from simulations (left) and observations (right). The properties shown are (a) the mass spectra, (b) cloud rotations and (c) the virial parameters of the clouds. The different symbols for the simulated clouds represent different surface density criteria used to select the clouds. For the observational plots, (a) and (c) are for clouds in the M51 galaxy, while (b) is from clouds in M33. Different galaxies were used as the required observational data to illustrate all three properties were not available for a single galaxy. Different galaxies, and simulations, will also exhibit different ranges of clouds masses. The observational data are from Colombo *et al.* (submitted) and Rosolowsky *et al.* (2003), the simulated clouds from Dobbs and Pringle (2013).

observations in order to infer both whether the simulations are including the appropriate physics, and to give credence to the ideas of cloud formation and evolution emerging from the simulations. Figure 3 shows the properties of simulated and observed clouds from Dobbs and Pringle (2013), Rosolowsky *et al.* (2003) and Colombo *et al.* (submitted). Unlike the observations, the simulated clouds are not computed from actual CO emission, rather the clouds are defined using a simple clump-finding algorithm. CO emission could be used, although it is worth bearing in mind that these simulations probably underproduce CO, as they do not resolve very dense gas.

Figure 3 shows the cloud mass spectra, virial parameters, and cloud rotations for the clouds in simulations and observations. The cloud mass spectra, which plots the number of clouds of different masses, generally has a slope of around −2 in both simulations and observations, although there tends to be some variation. A slope of <−2 implies that most of the gas is in the more massive ($10^5$–$10^6$ $M_\odot$) GMCs. The virial parameter, α, is the ratio of gravitational to kinetic energy in a cloud. Generally α < 1 implies that the cloud is gravitationally bound (i.e. the cloud will collapse), and α > 2 implies the cloud is gravitationally unbound, (i.e. the cloud will expand, or split apart). For 1 < α < 2 the cloud can be either bound or unbound. One of the fundamental questions regarding molecular clouds is whether they are gravitationally bound and, if so, are the clouds collapsing or are they supported (e.g. by turbulence or magnetic fields)? Both the simulations and recent observations indicate a clear presence of unbound clouds, challenging the more traditional picture of quasi-static gravitationally bound clouds. This is not surprising for, as shown in figure 2, the evolution of GMCs is clearly very complex, some splitting into other clouds while others merge into larger clouds. This picture is consistent with more transient clouds. Figure 3 also shows that although there are unbound (and likewise bound) clouds, α does not tend to reach values strongly deviating from 1, e.g. >10 or <0.1. This is again unsurprising as strongly unbound clouds would be insufficiently dense to be detected or form stars, whereas strongly bound clouds would collapse and turn all their gas into stars, which is contradictory to the general view that only a fraction of the gas in clouds forms stars otherwise the star formation rate would be too high. It is worth noting though that the values of α are likely to depend on the tracer used. For example, H I clouds would be lower density and likely have higher values of α, whereas species that trace higher densities, and internal cloud structure, would likely trace dense, bound gas and exhibit lower values of α.

The lower panel of figure 3 shows the angular momenta of the clouds in simulations and observations. The angular momenta, $L_z$, measures the rotation of the clouds. The plots here show the absolute values of $L_z$, but both the simulations and observations find positive and negative values of $L_z$. Thus clouds exhibit both prograde and retrograde rotation with respect to the galaxy (here $L_z$ represents the innate rotation of the clouds, rather than the orbit of the clouds round the galaxy). The presence of retrograde rotating clouds is a surprising result, as we would expect vorticity to be conserved and therefore observe clouds rotating in the same sense as the rotation of the galaxy. However, the collision of one cloud with another causes the rotation to change. In the simulations, these collisions lead to a much more even distribution of pro and retrograde cloud rotations, and are a result of an inhomogenous, clumpy medium. In contrast, where cloud collisions are rare or ineffective (e.g. in a diffuse homogeneous medium), retrograde clouds are rare or non-existent. Thus cloud–cloud collisions can naturally explain the distribution of cloud angular momenta (an alternative explanation is magnetic braking, although this predominantly decreases cloud rotation, rather than changing the sign of the rotation).

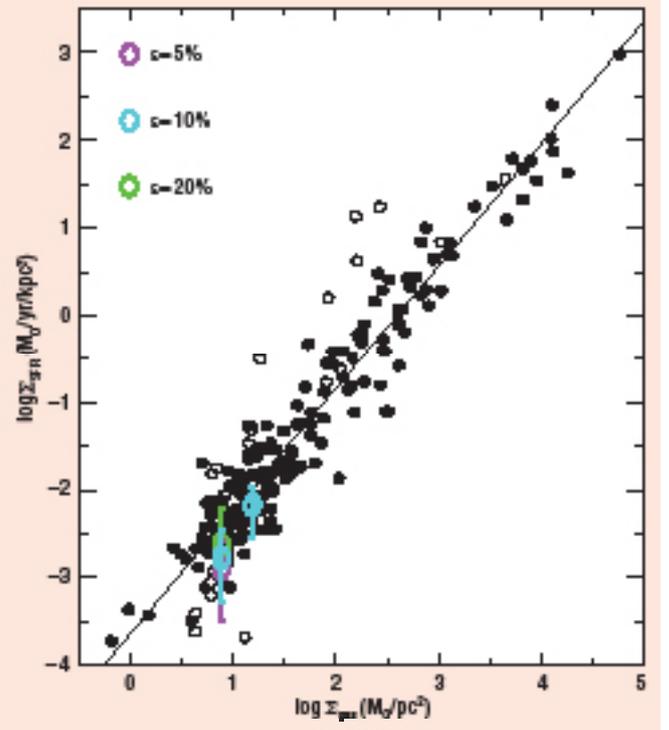

4: The star formation rate is plotted versus surface density for simulations (coloured points) with different levels of stellar feedback, and observed galaxies (filled black points are luminous spirals and irregulars, and open black points are fainter spiral and irregular galaxies).

## Star formation rates

As well as the properties of molecular clouds, another testable prediction of these simulations is the star formation rate in galaxies. Star formation in galaxies is particularly inefficient, the rate of star formation being much less than predicted from the amount of dense gas (Zuckerman and Evans 1974). In the particular simulations presented here, star particles are not included, although we do record when, where and how many stars form, each time a stellar feedback event occurs. Figure 4 shows the star formation rate versus surface density from simulations from Dobbs *et al.* (2011), with different levels of stellar feedback, plotted against observations from Kennicutt (2008). The observational trend of star formation rate versus surface density is more commonly known as the Schmidt–Kennicutt relation (Schmidt 1959, Kennicutt 1989). The Schmidt–Kennicutt relation has a slope of about $\Sigma_{SFR} \propto \Sigma^{1.4}$, although at lower density regimes there is a sharper drop off.

Figure 4 shows that the simulations fit the observed data closely, although there is considerable spread in the observed points. Interestingly, the level of feedback, as indicated by the efficiency parameter (ε) does not strongly affect the star formation rate. If the star formation efficiency parameter is doubled, twice as many stars form at each star formation event, but also the amount of energy added to the ISM is twice as large. However, the global star formation does not increase by as much as two-fold, indicating that fewer star formation events are occurring. Thus the star formation is effectively regulated by the stellar feedback.

The regulation by stellar feedback can be thought of as follows: if the star formation rate decreases, then there is less feedback. This means less energy is deposited in the ISM, and the gas becomes more gravitationally bound and more star formation occurs. The converse happens if the star formation rate increases. It is likely that stellar feedback (including winds and radiation pressure which are immediate, as well as supernovae which only occur at the end of a massive star's lifetime) regulates the virial parameters of clouds as well.

The existence of little or no feedback has a

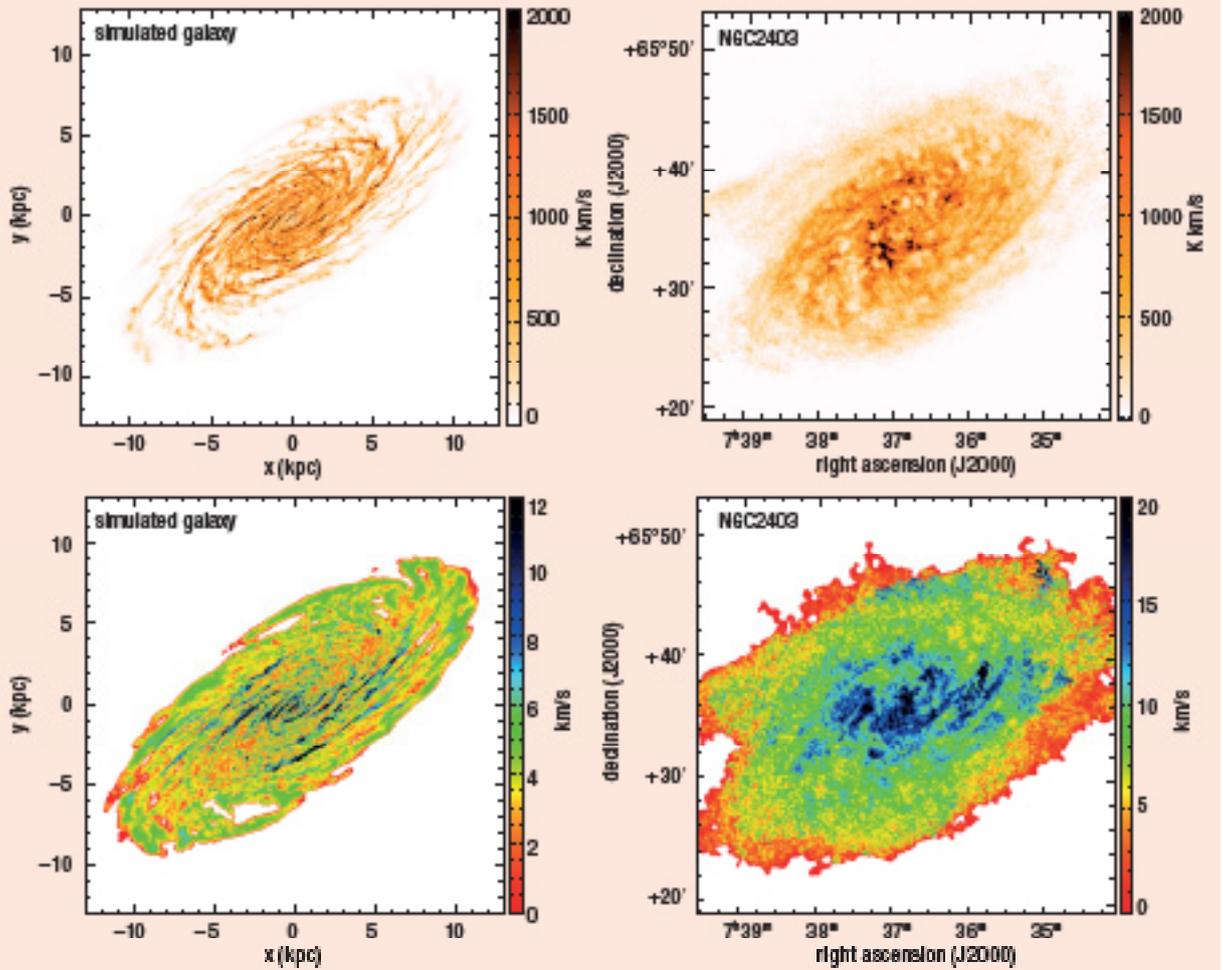

5: Synthetic maps of a simulated galaxy (left) are compared with the actual galaxy NGC2403 (right). The top panels show the integrated HI maps, while the bottom panels show the HI velocity dispersion. These figures are provided by David Acreman, from work by Acreman and Mphys students Freya Aldred, William English and Charlotte Harrison. The data for NGC2043 is taken from the THINGS survey (Walter *et al.* 2008).

very dramatic effect on the star formation rate. Without feedback, the star formation rate is very high, much higher than the observed Schmidt–Kennicutt relation. In Dobbs *et al.* (2011), a galaxy simulation with minimal feedback formed so many stars, and the clouds became so gravitationally dominated, that the calculation became very slow and impractical to run further. Simulations of local clouds (Bonnell *et al.* 2013, Van Loo *et al.* 2013) also find that without feedback, the star formation rates are much higher than those observed. Other simulations of galaxies and local molecular clouds also confirm the need for stellar feedback to obtain realistic star formation rates (Vázquez-Semadeni *et al.* 2010, Hopkins *et al.* 2011).

In fact the regulation of star formation by feedback also appears vital to producing realistic disc galaxies. In the past, galaxy formation simulations tended to produce bulge-dominated galaxies. However, simulations with more realistic feedback implementations, and which add feedback at more realistic gas densities, produce more spiral-like disc galaxies, with rotation curves more similar to observations (Stinson *et al.* 2013, Agertz *et al.* 2011).

### Synthetic observations

In order to truly compare simulations and observations, we ideally need to convert the output from the simulations into maps of emission using the same tracers as used by observers, then use the same tools (e.g. clump-finding algorithms, dendrograms) as observers to study molecular clouds and the structure of the ISM. This area of work is still in its relative infancy, but synthetic HI (atomic hydrogen) maps have been produced from the observations (Acreman *et al.* 2010, 2012). To produce the HI maps, the SPH simulations are post-processed using the TORUS radiative transfer code (Harries 2011).

The synthetic emission maps can be compared to both galactic (e.g. Canadian Galactic Plane Survey, CGPS) and extragalactic HI surveys (e.g. The HI Nearby Galaxy Survey, THINGS). Properties of the ISM that can be calculated from the HI maps include the velocity dispersion, scale height of the disc and distribution/size of supernova shells. These properties help determine which physics is important in producing the structure and dynamics of the ISM. For example, Acreman *et al.* (2012) demonstrated that stellar feedback is required to produce a realistic scale height of the HI disc compared with our galaxy. Figure 5 shows synthetic maps for a simulated flocculent (many-armed) galaxy compared to the real flocculent spiral galaxy NGC2403. The figure shows maps of both the HI emission and the velocity dispersion in the HI gas. The velocity dispersion is a measure of the random component of the velocity field and in the simulations is largely determined by stellar feedback. The agreement of the values of HI emission and velocity dispersion obtained in the simulated and real galaxy are encouraging, and suggest that the basic properties of the galaxies we are simulating are correct. However, there are clear differences between the simulated and actual galaxy. The arms are less clear and wider in NGC2403. This could partly reflect lower resolution of the HI observations, but may also suggest that some physics is absent from the simulation (e.g. magnetic fields, a bar, tidal interactions) or perhaps that an aspect of the included physics could be improved (e.g. stellar feedback).

CO traces the molecular gas and thus is the best way to study molecular clouds but, so far, the study of CO formation has largely been confined to smaller scale simulations of molecular clouds, e.g. simulations of turbulent boxes or colliding flows (Shetty *et al.* 2011, Heitsch and Hartmann 2008). Synthetic observations can also be used to probe the spiral structure of our galaxy by comparing maps of HI and CO with those observed (Rodriguez-Fernadez and Combes 2008, Pettitt *et al.* 2013).

### From galaxies to star formation

While informative about the formation and evolution of GMCs, these simulations of galaxies are still far removed from the scales of individual star formation. One way to reach smaller scales is to resimulate a section of a global galaxy simulation, and effectively zoom in on star formation. This has now been achieved for the first time by a couple of groups, Bonnell *et al.* (2013) and Van Loo *et al.* (2013).

Figure 6 is taken from Bonnell *et al.* (2013) and illustrates three different simulations. The first is a global galaxy simulation (modelling just a torus of the galaxy to maximize resolution). A section of this simulation is then resimulated (for a shorter time scale) at higher resolution. Then a third resimulation is carried out, zooming in on a region of the second simulation. By resimulating progressively smaller scales, the resolution of the third simulation reaches a particle mass of $0.156\,M_\odot$. This is still not quite enough to be able to follow the formation of individual, $1\,M_\odot$ stars (to follow star formation, we typically insert sink particles to denote stars that have formed; in SPH codes, a sink particle typically replaces 50–100 SPH particles). Instead the minimum mass that can be resolved is $11\,M_\odot$, so the insertion of sink particles in the simulation represents either massive stars, or a small group of stars.

These high-resolution simulations provide a better handle on what determines the Schmidt–Kennicutt relation on cloud scales, and the nature of star formation within a cloud. In particular, Bonnell *et al.* (2013) found that cooling and shocks have a strong role in governing the amount of cold, dense gas. They find that it is these processes, which are largely associated with the larger scale dynamics of the ISM, that drive the Schmidt–Kennicutt relation. There is a nonlinear relation between the amount of cold gas and the total amount of gas, but then the relation between cold gas and molecular gas is linear. However, without any stellar feedback, or magnetic fields in these simulations, the star formation rates are found to be too high compared to those observed. Thus stellar feedback, and possibly magnetic fields, are seemingly required to obtain the correct normalization of the Schmidt–Kennicutt relation.

In the future, we aim to have a much more complete picture of how star formation progresses from the scale of individual stars to galactic scales, incorporating large-scale processes such as spiral arms, supernovae and galactic rotation. Notably, stellar feedback has a clear role in regulating star formation, depositing energy in the ISM, and contributing to the structure of molecular clouds. However, the role of magnetic fields and their relevance to cloud formation, cloud dispersal and regulating star formation has not yet been investigated. Furthermore, spiral galaxies should ideally be modelled in a more realistic context, including the generation of spiral arms, but naturally including extra physics, or larger scales, makes achieving the high resolution required to model GMCs challenging. ●

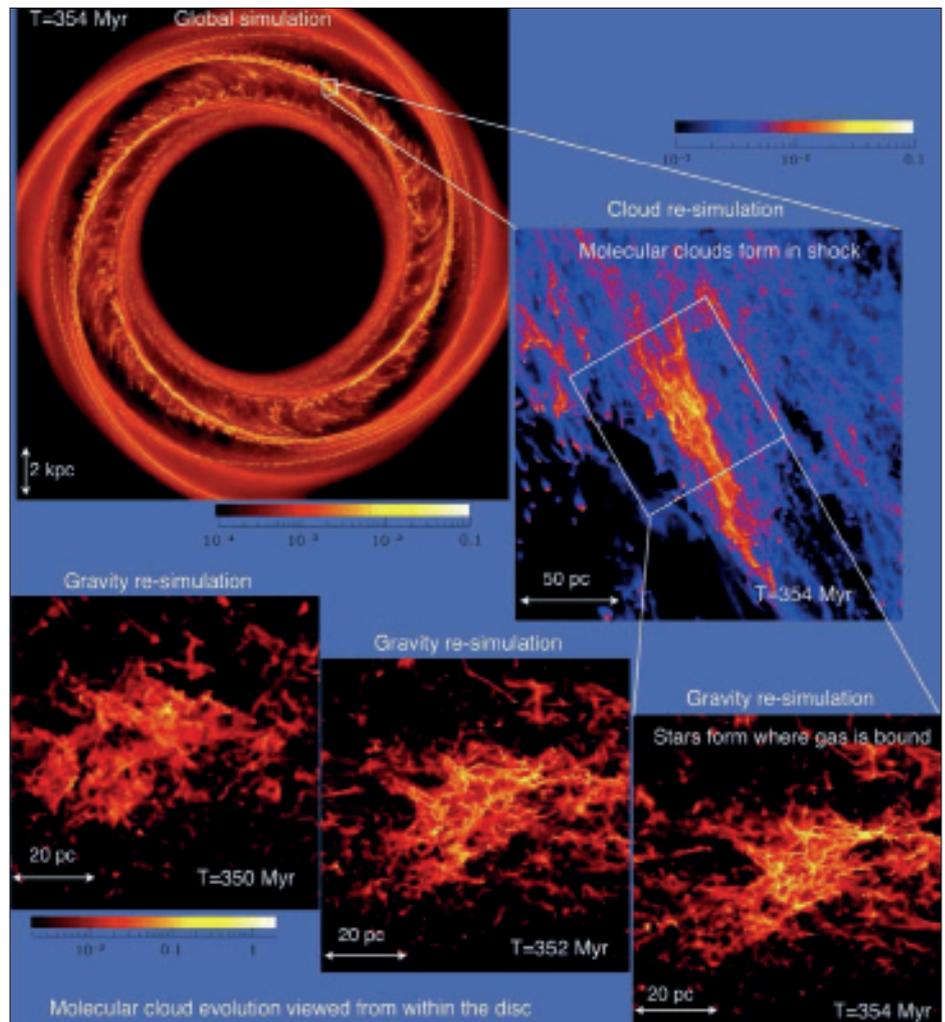

**6:** A schematic of three simulations which zoom in on a region of star formation in a galaxy, in order to study in detail how a star forming region is formed in a galaxy, and how star formation progresses on parsec to sub-parsec scales (Bonnell *et al.* 2013).

*Clare Dobbs is a proleptic lecturer at the School of Physics & Astronomy, University of Exeter, UK; dobbs@astro.ex.ac.uk.*

*Acknowledgments. I would like to thank David Acreman and Ian Bonnell for providing figures 4 and 6, and Jim Pringle, with whom I collaborated on much of this work. This research is funded by the ERC starting grant project LOCALSTAR, and uses the DIRAC national HPC facilities.*

**References**
**Acreman D M et al.** 2010 *Mon. Not. Roy. Astron. Soc.* **406** 1460.
**Acreman D M et al.** 2012 *Mon. Not. Roy. Astron. Soc.* **422** 241.
**Agertz O et al.** 2011 *Mon. Not. Roy. Astron. Soc.* **410** 1391.
**Bate M** 1995 PhD thesis University of Cambridge.
**Benz W et al.** 1990 *Ap. J.* **348** 647.
**Blitz L and Shu F H** 1980 *Ap. J.* **238** 148.
**Bonnell I A et al.** 2013 *Mon. Not. Roy. Astron. Soc.* **430** 1790.
**Dobbs C L** 2008 *Mon. Not. Roy. Astron. Soc.* **391** 844.
**Dobbs C L and Pringle J E** 2013 *Mon. Not. Roy. Astron. Soc.* **432** 653.
**Dobbs C L et al.** 2011 *Mon. Not. Roy. Astron. Soc.* **417** 1318.
**Goldreich P and Lynden-Bell D** 1965 *Mon. Not. Roy. Astron. Soc.* **130** 125.
**Harries T J** 2011 *Mon. Not. Roy. Astron. Soc.* **416** 1500.
**Heitsch F and Hartmann L** 2008 *Ap. J.* **689** 290.
**Hopkins P F et al.** 2011 *Mon. Not. Roy. Astron. Soc.* **417** 950.
**Kennicutt R C** 1989 *Ap. J.* **344** 685.
**Kennicutt R C** 2008 The Schmidt Law: Is it Universal and What Are its Implications? in *Pathways Through an Eclectic Universe 2008* Astronomical Society of the Pacific Conference Series 390 eds J H Knapen *et al.* 149.
**Kim W-T et al.** 2002 *Ap. J.* **581** 1080.
**La Vigne M A et al.** 2006 *Ap. J.* **650** 818.
**Pettitt A R et al.** 2013 in *Proceedings of IAUS* **298** eds Röllig M *et al.*
**Price D J and Monaghan J J** 2007 *Mon. Not. Roy. Astron. Soc.* **374** 1347.
**Rodriguez-Fernandez N J and Combes F** 2008 *A. & A.* **489** 115.
**Rosolowsky E et al.** 2003 *Ap. J.* **599** 258.
**Schmidt M** 1959 *Ap. J.* **129** 243.
**Shetty R and Ostriker E C** 2006 *Ap. J.* **647** 997.
**Shetty R et al.** 2011 *Mon. Not. Roy. Astron. Soc.* **412** 1686.
**Stinson G S et al.** 2013 *Mon. Not. Roy. Astron. Soc.* **428** 129.
**Toomre A** 1964 *Ap. J.* **139** 1217.
**Van Loo S et al.** 2013 *Ap. J.* **764** 36.
**Vázquez-Semadeni E et al.** 2010 *Ap. J.* **715** 1302.
**Wada K and Koda J** 2004 *Mon. Not. Roy. Astron. Soc.* **349** 270.
**Wada K and Norman C A** 1999 *Ap. J. Letts* **516** L13.
**Walter F et al.** 2008 *Astron. J.* **136** 2563.
**Zuckerman B and Evans II N J** 1974 *Ap. J.* **192** L149.